# Improved modeling of dynamic quantum systems using exact Lindblad master equations

Jacob R. Lindale[1], Shannon L. Eriksson[1,2], Warren S. Warren[1,3]

[1] Department of Chemistry, Duke University, Durham, NC, 27705. [2] School of Medicine, Duke University, Durham, NC, 27705.

[3] Departments of Physics, Biomedical Engineering, & Radiology, Duke University, Durham, NC, 27705.

The theoretical description of the interplay between coherent evolution and chemical exchange, originally developed for magnetic resonance and later applied to other spectroscopic regimes, was derived under incorrect statistical assumptions. Correcting these assumptions provides access to the exact form of the exchange interaction, which we derive within the Lindblad master equation formalism for generality. The exact form of the interaction is only different from the traditional equation by a scalar correction factor derived from higher-order interactions and regularly improves the radius of convergence of the solution (hence increasing the allowable step size in calculations) by up to an order of magnitude for no additional computational cost.

## I. Introduction

Chemical exchange encompasses a broad scope of molecular dynamics that results in a change of the system Hamiltonian and may be interrogated spectroscopically. Techniques such as nuclear magnetic resonance (NMR), two-dimensional infrared spectroscopy (2DIR), and chirped pulse Fourier Transform microwave spectroscopy (CP-FTMS), are sensitive to exchange processes on vastly different timescales[1-10]. For example, these techniques have been used to study the *in-situ* structure and functionality of biomolecules, hydrogen bonding dynamics in liquids, and dynamic rotational isomerization. The theoretical treatment presented here is also applicable to discrete multisite exchange problems, such as atoms migrating in an optical trap array. Interpreting the experimental data from any of these techniques often requires the use of a physical model that unifies the coherent and chemical dynamics.

The density matrix formalism is a convenient method to include statistical averaging in systems that evolve coherently and is ubiquitous in spectroscopy. Kaplan[11, 12] (1958) and Alexander[13, 14] (1962) were the first to describe the exchange interaction within the density matrix formalism, motivating the form of the interaction from first principles to describe the NMR lineshape under exchange. In essence, they described exchange by the transformation

$$\hat{\rho} \to \hat{R}\hat{\rho}\hat{R}^{-1} - \hat{\rho}, \qquad (1)$$

where $\hat{R}\hat{\rho}\hat{R}^{-1}$ was a similarity transform relating the density matrix before and after an exchange event. For this reason, exchange is often discussed as passing population between different "sites", where each site has a unique molecular geometry and associated lifetime. Binsch[15] (1969) later unified Kaplan and Alexander's theory of chemical exchange with Redfield's relaxation theory[16] (1957) to fully describe the NMR lineshape. By the advent of coherent optical spectroscopies, the traditional form of the exchange interaction had been supported by decades of experimental evidence from the magnetic resonance community and was thus adopted by the ultrafast spectroscopy community.

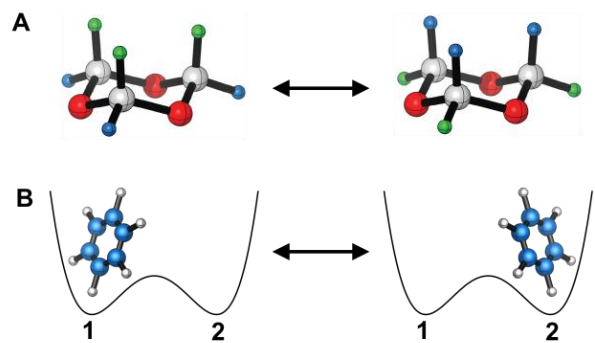

**Figure 1 | Manifestations of exchange. A.** Chemical exchange often results in the rearrangement of molecules, which may be both intramolecular (shown here) or intermolecular. $^1$H nuclei are colored to distinguish configurations. **B.** Molecular rearrangement is not necessarily required if coherent interactions are site dependent and the species interchanges between sites 1 and 2.

Recently, the theoretical underpinnings of Binsch's unification of quantum dynamics and chemical exchange were found to be flawed, because it motivated that the exchange interaction was simply an extension of the Redfield's relaxation[17]. Thermalization in liquid state magnetic resonance is generated by modulation of orientationally-dependent interactions during molecular dynamics on the fs-ps timescale. Fourier components of this modulation on the order of NMR transitions then induces transitions that drive the system towards equilibrium. However, the molecular processes that are described by Redfield's theory and chemical exchange dynamics are different. While molecular tumbling is continuous, chemical exchange is a fundamentally discontinuous process, as the time required for molecular rearrangement is often orders of magnitude faster than the lifetime in any particular site. These two cases require different statistical assumptions that directly impact the formulation of the chemical exchange interaction.

Recently, we introduced an exact dissipative master equation[17] (DMEx) for chemical exchange that was rigorously derived as a closed form solution of the Dyson expansion. This work agreed with Binsch's treatment, under more justified assumptions, but only to lowest order in perturbation theory. It was derived using projection or pseudorotation superoperators, which are both common in magnetic resonance. The resulting differential equation was proportional to a traditional exchange term by a scalar factor, which was obtained by contraction of the higher order terms of the Dyson series. Including this correction improves the convergence radius of the series by up to an order of magnitude for no additional computational cost. However, the formulation using projection and pseudorotation superoperators restricts the generality of the formalism.

Here, we introduce a generalization of the exact Dissipative Master equation formalism and extend the derivation to encompass both the cases of an arbitrary exchange pathway as well as the case where the exchange pathway has permutation group symmetry. For compatibility, we shall treat exchange as a Lindblad[18-20] equation (also referred to as the Gorini-Kossakowski-Sudarshan-Lindblad equation), as it is the most general Markovian master equation. We will show that an exact Lindblad master equation (LMEx) can be derived by continuing the traditional derivation to infinite-order in the Dyson series. Within this formalism, we will show that the exchange interaction in its most general form can be written as:

$$\hat{\mathcal{L}}\hat{\rho} = \frac{1}{\tau}\left(\hat{A}\hat{\rho}\hat{A}^\dagger - \frac{1}{2}\{\hat{A}\hat{A}^\dagger, \hat{\rho}\}\right)\exp\left(\frac{-T}{\tau}\right)$$

This is isomorphic to **eq. 1**, where $\hat{A}$ is a Hermitian operator analogous to $\hat{R}$ and imposes exchange on the system at a rate $1/\tau$. The only difference from this equation and the previous formalism for chemical exchange is the exponential factor $\exp(-T/\tau)$ where $T$ is the time over which the master equation is averaged (usually, the time step in a calculation). This factor is derived as the closed form solution of the Dyson series and has the practical implication of increasing the radius of convergence up to an order of magnitude over the traditional master equation at no additional computational cost.

## II. Derivation of a generalized Lindblad master equation for chemical exchange

We will show that the exact form of the exchange interaction may be derived *ab initio* using a minimal number of foundational assumptions. To begin, we will establish the assumptions to be used throughout the derivation:

(1) Exchange is a Hermitian, multiparticle coupling, and therefore has a corresponding Hamiltonian $\hat{\mathcal{H}}_1(t)$.
(2) The system is Markovian, permitting us to make the substitution $\hat{\rho}(t_n) \to \hat{\rho}(t)$, where $t > t_n$ by time-ordering.
(3) To satisfy the Hermiticity requirement, the system is at a steady state. As such, the dynamics at the level of the ensemble are assumed to be stationary.
(4) The time required for molecular rearrangement is much faster than any other interaction in the system and may thus be assumed to be instantaneous.

While not an assumption, we will stipulate that any exchange process is not correlated to any other process. To satisfy this requirement, the basis set of exchange processes are transformed such that they are orthogonal, thus automatically satisfying this requirement. This will facilitate deriving the higher-order Dyson series terms.

It is important to note that these assumptions are identical to those established for the DMEx. The substantial difference between this treatment and the DMEx is that the form of the exchange Lindbladian is determined *a priori*, as opposed to the exchange superoperator, which is less stringently defined. We will show that the exchange Lindbladian spans a well-defined composite Hilbert-Fock space and only acts on the Fock state, whereas the exchange superoperator acts directly on the Hilbert state of the system. As such, the DMEx and Lindblad treatments of exchange are complimentary, and are suited for different purposes. Direct action on the Hilbert-space is ideal for cases when molecules dissociate and the dimensionality of the Hilbert-space changes. The action of the Lindbladian on the Fock-space makes this formulation significantly more general.

At this juncture, we may begin deriving the Lindblad master equation for chemical exchange, which treats the entire system quantum mechanically before reducing the density matrix. Under assumption (1), that exchange is Hermitian and has a Hamiltonian, we may define the system Hamiltonian as:

$$\hat{\mathcal{H}}(t) = \hat{\mathcal{H}}_0 + \hat{\mathcal{H}}_1(t) \quad (2)$$

We have partitioned this into a static component, $\hat{\mathcal{H}}_0$ that contains the coherent interactions, and a stochastically modulated component $\hat{\mathcal{H}}_1(t)$ describing the exchange contribution to the Hamiltonian. Using this within the Liouville-von Neumann equation gives ($\hbar = 1$):

$$\frac{\partial}{\partial t}\hat{\rho}(t) = -i[\hat{\mathcal{H}}_0 + \hat{\mathcal{H}}_1(t), \hat{\rho}(t)] \quad (3)$$

This may be simplified by transforming into the interaction representation (dropping hats to denote frame) as:

$$\frac{\partial}{\partial t}\rho(t) = -i[\mathcal{H}_1(t), \rho(t)] \quad (4)$$

Formally integrating this result gives:

$$\rho(t) = \rho_0 - i\int_0^t dt' \, [\mathcal{H}_1(t'), \rho(t')] \quad (5)$$

**Equation 5** may be iteratively substituted into **eq. 4** to give the terms of the Dyson series, where the first two terms are:

$$\frac{\partial}{\partial t}\rho^{(1)} = -i[\mathcal{H}_1(t), \rho(t)] \quad (6)$$

$$\frac{\partial}{\partial t}\rho^{(2)} = -\vec{\mathcal{T}}\int_0^t dt_1 \, \left[\mathcal{H}_1(t), [\mathcal{H}_1^\dagger(t_1), \rho(t_1)]\right] \quad (7)$$

$\vec{\mathcal{T}}$ is the Dyson time-ordering operator that enforces $t > t_1$. Note that we have indicated the term of the Dyson series by $\rho^{(n)}$ and have dropped the formal time dependence on the left-hand side for brevity. Before continuing, we introduce an operator expansion of $\mathcal{H}_1(t)$ for chemical exchange (**Fig. 2**) as the tensor product between an operator $A_k$ that acts on the Fock space of the system to generate exchange and the stochastically modulated operator $F_k$ that describes the molecular dynamics:

$$\mathcal{H}_1(t) = \sum_k A_k(t) \otimes F_k(t) \quad (8)$$

Ensemble averaging only affects the $F_k$ operators, as they carry the stochastic modulation. In accordance with the assumption that the ensemble dynamics are stationary, we define the operator $\hat{A}_k$ in terms of Fock-space creation ($\hat{a}_k^\dagger$) and annihilation ($\hat{a}_k$) operators that generate transitions between two sites connected by an exchange process as:

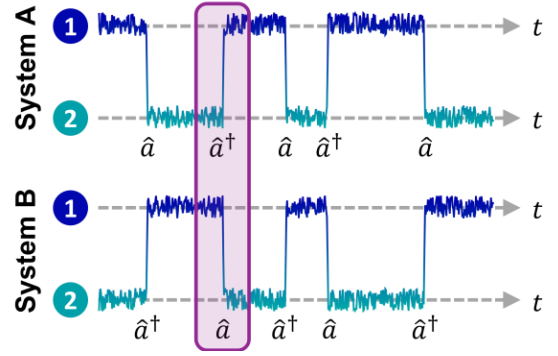

**Figure 2 | Formulation of the exchange Hamiltonian.** For the ensemble to be stationary, two systems (A and B) must simultaneously exchange between sites 1 and 2. As such, we impose the pairwise action of the $\hat{a}$ and $\hat{a}^\dagger$, which leads to the definition of $\hat{A}$. The operator $\hat{F}(t)$ describes the molecular trajectories of each dynamic process.

$$\hat{A}_k = \hat{a}_k + \hat{a}_k^\dagger \quad (9)$$

As $\hat{F}_k(t)$ is stochastic and varies for each member of the ensemble, we will only be able to define statistical metrics of this operator over the entire ensemble.

The Markovian assumption (2) permits the substitution $\rho(t_1) \approx \rho(t)$ in each term of the Dyson series. Furthermore, the stationary assumption (3) has the ramification that $\langle F_k(t) \rangle = 0$, as this term would generate drift in the stochastic motion. As such, the leading observable term is the second order term from the expansion. The only time-parameter of importance is the difference between $t$ and $t_1$, and can use the change of variables $t - t_1 = \tau$ to give:

$$\frac{\partial}{\partial t}\rho^{(2)} = \vec{\mathcal{T}}\sum_{jk}\int_t^0 d\tau \begin{bmatrix} A_j(\tau) \otimes F_j(\tau), \\ [\hat{A}_k^\dagger \otimes \hat{F}_k^\dagger, \rho] \end{bmatrix} \quad (10)$$

Note that the limits of integration are swapped with this transformation and that the differential transforms as $dt_1 \to -d\tau$. In doing this, the operators $\hat{A}_k$ and $\hat{F}_k$ lose time-dependence and thus can be written in the Schrödinger representation (with hats). At this juncture, it is pertinent to isolate the degrees of freedom in $\rho$ corresponding to the stochastic operators, which is delineated as $\rho_B$. Inverting the limits of integration, expanding the double commutator, and averaging over the stochastic components gives ($\vec{\mathcal{T}}$ is assumed):

$$\frac{\partial}{\partial t}\rho^{(2)} = \sum_{jk}\int_0^t d\tau \begin{pmatrix} A_j(\tau)\rho\hat{A}_k^\dagger \\ -A_j(\tau)\hat{A}_k^\dagger\rho \end{pmatrix}\langle F_j(\tau)\hat{F}_k^\dagger\rho_B\rangle$$
$$+ \int_0^t d\tau \begin{pmatrix} \hat{A}_k^\dagger\rho A_j(\tau) \\ -\rho\hat{A}_k^\dagger A_j(\tau) \end{pmatrix}\langle \hat{F}_k^\dagger F_j(\tau)\rho_B\rangle \quad (11)$$

Here, the angle brackets indicate the partial trace over the stochastic degrees of freedom. We shall assume that at the level of the ensemble, the stochastically distributed exchange events are independent of one another and uniformly distributed, which permits us to assume that exchange is a Gaussian process. Doing so permits us to rearrange the correlation function under Isserlis' theorem as:

$$\langle F_j(\tau)\hat{F}_k^\dagger \rho_B \rangle = \langle F_j(\tau)\hat{F}_k^\dagger \rangle \langle \rho_B \rangle \\ + \langle F_j(\tau)\rho_B \rangle \langle \hat{F}_k^\dagger \rangle + \langle \hat{F}_k^\dagger \rho_B \rangle \langle F_j(\tau) \rangle \quad (12)$$

Due to the stationary assumption, only the first term survives averaging. Furthermore, $\langle \rho_B \rangle = \hat{E}$ (the identity matrix) by definition. It is now pertinent to introduce the form of the correlation functions. For chemical exchange, the correlation function $\langle F_j(\tau)\hat{F}_k^\dagger \rangle$ dictates both the rate of exchange as well as the jump time between sites, which under assumption (4) is instantaneous, or $\delta$-correlated in time. Finally, our stipulation that exchange processes are only self-correlated imposes quadratic action of any term in the exchange Hamiltonian, which is required to exchange population between sites. As such, we may write the correlation functions as:

$$\langle F_j(\tau)\hat{F}_k^\dagger \rangle = \frac{\delta(\tau)}{\tau_k}\delta_{jk} \quad (13)$$

$\tau_k$ is the characteristic lifetime in the two configurations connected by the process $\hat{F}_k$. **Equations 12** and **13** permit us to extend the integration limits and rewrite **eq. 9** as:

$$\frac{\partial}{\partial t}\rho^{(2)} = \sum_k \frac{1}{2}\int_{-\infty}^{\infty} d\tau \begin{pmatrix} A_k(\tau)\rho\hat{A}_k^\dagger \\ -A_k(\tau)\hat{A}_k^\dagger\rho \end{pmatrix}\frac{\delta(\tau)}{\tau_k} \\ + \frac{1}{2}\int_{-\infty}^{\infty} d\tau \begin{pmatrix} \hat{A}_k^\dagger \rho A_k(\tau) \\ -\rho\hat{A}_k^\dagger A_k(\tau) \end{pmatrix}\frac{\delta(\tau)}{\tau_k} \quad (14)$$

Note that we have acquired a factor of 1/2 by also taking the lower integration limit to $-\infty$ such that the $\delta(\tau)$ function is real-valued. Performing integration greatly simplifies the expression to:

$$\frac{\partial}{\partial t}\rho^{(2)} = \sum_k \frac{1}{2\tau_k}\begin{pmatrix} \hat{A}_k\rho\hat{A}_k^\dagger + \hat{A}_k^\dagger\rho\hat{A}_k \\ -\hat{A}_k\hat{A}_k^\dagger\rho + \rho\hat{A}_k^\dagger\hat{A}_k \end{pmatrix} \quad (15)$$

Remembering that $\hat{A}_k$ are Hermitian, we may cast this into the conventional Lindblad form, using $\{\hat{A},\hat{B}\} = \hat{A}\hat{B} + \hat{B}\hat{A}$ as the notation for the anticommutator:

$$\frac{\partial}{\partial t}\rho^{(2)} = \sum_k \frac{1}{\tau_k}\left(\hat{A}_k\rho\hat{A}_k^\dagger - \frac{1}{2}\{\hat{A}_k\hat{A}_k^\dagger,\rho\}\right) \quad (16)$$

This is the traditional master equation for chemical exchange written in Lindblad form. The term $\hat{A}_k\rho\hat{A}_k^\dagger$ generates the similarity transform relating the two sites and the term proportional to the anticommutator simply reduces to $\hat{\rho}$ when the $\hat{A}_k$ operators interchange populations between two of the Fock states. The entire equation is isomorphic to the form of the exchange interaction originally motivated by Kaplan and Alexander. This equation can be recovered if one begins from the traditional form of the exchange interaction[21], however the benefit of deriving the result *ab initio* is that the framework is established to calculate higher-order interactions. It will be advantageous write **eq. 16** as

$$\frac{\partial}{\partial t}\rho^{(2)} = \sum_k \frac{\hat{\mathcal{L}}_k\rho}{\tau_k}, \quad (17)$$

where $\hat{\mathcal{L}}_k$ is the Lindbladian dissipator corresponding to the term in parentheses. For brevity, we will call this the second-order Lindblad master equation (LME2), which is the exchange analog of the conventional Lindblad master equation derived for open quantum systems.

Traditionally, the exchange interaction is assumed to be small relative to all other interactions, which allows the Dyson series to be truncated to its leading term. While higher order exchange interactions become irrelevant if **eq. 14** can be analytically integrated, it is rarely possible to accomplish this in realistic systems. Instead, it is beneficial to think of this result in terms of a numerical simulation method. The LME2 will no longer be representative of the physical system as soon as the integration grid approaches a spacing where two or more exchange events become possible. As such, convergence of the simulation requires integration grids that are finely spaced so to prevent multiple exchange events.

We may continue the derivation of higher order exchange interactions by substituting **eq. 8** into the fourth-order term of the Dyson series, noting that the stationary assumption only permits even-order terms to be non-zero:

$$\frac{\partial}{\partial t}\rho^{(4)} \\ = \vec{\mathcal{T}}\sum_{jklm}\iiint_0^t dt_n \langle F_j(t)F_k^\dagger(t_1)F_l(t_2)F_m^\dagger(t_3)\rho_B \rangle \\ \times \left[A_j(t),\left[A_k^\dagger(t_1),\left[A_l(t_2),\left[A_m^\dagger(t_3),\rho\right]\right]\right]\right] \quad (18)$$

The restriction that exchange processes are only self-correlated processes permits arbitrary re-indexing of the operators, allowing the four-point correlator to be factored out of the commutators and averaged. We have used:

$$\vec{\mathcal{T}} \iiint_0^t dt_n \cdots = \vec{\mathcal{T}} \int_0^t dt_1 \int_0^{t_1} dt_2 \int_0^{t_2} dt_3 \cdots \quad (19)$$

to represent the time-ordered integral. Using Isserlis' theorem, we may also note that the only terms that will be non-zero will be those where $\rho_B$ is averaged separately from the stochastic operators, similar to **eq. 12**. This allows us to write the four-point correlation function as:

$$\begin{aligned}&\langle F_j(t) F_k^\dagger(t_1) F_l(t_2) F_m^\dagger(t_3) \rho_B \rangle \\&= \langle F_j(t) F_k^\dagger(t_1) \rangle \langle F_l(t_2) F_m^\dagger(t_3) \rangle \\&+ \langle F_j(t) F_l(t_2) \rangle \langle F_k^\dagger(t_1) F_m^\dagger(t_3) \rangle \\&+ \langle F_j(t) F_m^\dagger(t_3) \rangle \langle F_k^\dagger(t_1) F_l(t_2) \rangle\end{aligned} \quad (20)$$

**Equation 13** allows us to define these quantities as:

$$\langle F_j(t) F_k^\dagger(t_1) F_l(t_2) F_m^\dagger(t_3) \rho_B \rangle \quad (21)$$

$$= \frac{1}{\tau_k^2} \begin{pmatrix} \delta(t-t_1)\delta(t_2-t_3) + \\ \delta(t-t_2)\delta(t_1-t_3) + \\ \delta(t-t_3)\delta(t_1-t_2) \end{pmatrix} \delta_{jklm}$$

$$= \frac{3}{\tau_k^2} \delta(t-t_1)\delta(t_2-t_3)\delta_{jklm}$$

The second equality is permitted as the only time at which **eq. 21** is non-zero is when all time variables are identical. Effectively, this allows us to freely re-index the time-variables.

The time-ordered integral will become cumbersome for successively higher-order terms, but can be simplified by uncoupling the integrals. This is accomplished by dividing the resulting equation where the integrals are uncoupled by the number of degenerate time orderings. Generally, for the $n^{\text{th}}$-order term of the Dyson series, the time-ordering degeneracy will be $(n-1)!$ upon integration. We will again use the stationary assumption to make the change of variables $\tau_1 = t - t_1$ and $\tau_2 = t_2 - t_3$. This, along with **eq. 21**, let us rewrite **eq. 18** as:

$$\frac{\partial}{\partial t} \rho^{(4)} = \sum_k \frac{1}{\tau_k^2} \frac{3}{3!} \left( \int_t^0 d\tau_1 \left[ A_k(\tau_1), \left[ \hat{A}_k^\dagger, \int_t^0 d\tau_2 \left[ A_k(\tau_2), [\hat{A}_k^\dagger, \rho] \right] \delta(\tau_2) \right] \right] \delta(\tau_1) \right) \int_0^t d\tau_2 \quad (22)$$

The limits of integration may be reversed for two of the integrals within the parentheses without obtaining a sign-change, and we will choose to extend the limits of integration for these two integrals to $\pm\infty$, each time acquiring a factor of ½. The final integral arises from the $dt_2$ integral after the change of variables, and can be effectively factored out of the commutators. For clarity, we will simply change the upper limit of integration from $t \to T$, as this limit effectively denotes the time over which the equation of motion is averaged. This gives:

$$\frac{\partial}{\partial t} \rho^{(4)} = \frac{1}{4} \sum_k \frac{1}{\tau_k^2} \frac{3}{3!} \left( \int_{-\infty}^\infty d\tau_1 \left[ A_k(\tau_1), \left[ \hat{A}_k^\dagger, \int_{-\infty}^\infty d\tau_2 \left[ A_k(\tau_2), [\hat{A}_k^\dagger, \rho] \right] \delta(\tau_2) \right] \right] \delta(\tau_1) \right) \int_0^T d\tau_2 \quad (23)$$

Performing integration and simplifying yields:

$$\frac{\partial}{\partial t} \rho^{(4)} = \sum_k \frac{1}{4} \left[ \hat{A}_k, \left[ \hat{A}_k^\dagger, \left[ \hat{A}_k, [\hat{A}_k^\dagger, \rho] \right] \right] \right] \frac{T}{2\tau_k^2} \quad (24)$$

We must now make a distinction on the type of system being studied (**Fig. 3**), in particular to identify if the exchange processes are distinguishable or indistinguishable. We will restrict ourselves to delineate these cases based on the time-scale over which the interaction is expected to be modeled. For instance, it is routine to measure magnetic resonance signals over several seconds, so only interactions that evolve over this time period can lead to an inequivalence between sites and be classified as a distinguishable process. For distinguishable exchange processes where $\hat{F}_j \neq \hat{F}_k$, the $\delta_{jklm}$ function holds and **eq. 24** may be written as:

$$\frac{\partial}{\partial t} \rho^{(4)} = \sum_k \left( \frac{\hat{L}_k}{\tau_k} \right)^2 \rho \frac{T}{2} \quad (25)$$

However, when the exchange processes are indistinguishable, such that $\hat{F}_j = \hat{F}_k$ after re-indexing, the $\delta_{jklm}$ function must be bifurcated into $\delta_{jk}\delta_{lm}$, which permits different exchange processes $\hat{A}_j$ and $\hat{A}_k$ to be coupled. Additionally, $\tau_j = \tau_k$ when $\hat{F}_j = \hat{F}_k$, which will simply be called $\tau$ for simplicity. As

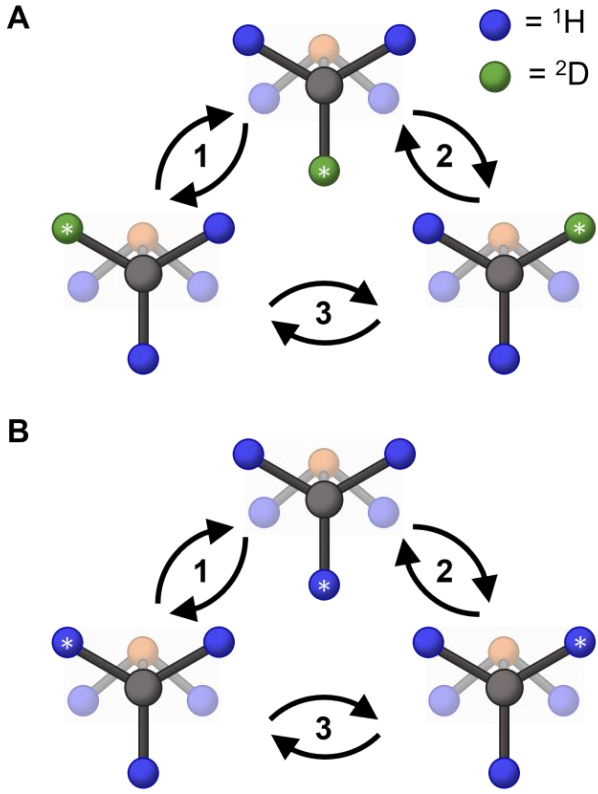

**Figure 3 | Distinguishable and indistinguishable exchange processes on methylamine isotopologues.** The ∗ indicates the position of exchanged atoms. **A.** In CH$_2$DNH$_2$, exchange processes 1 and 2 are distinguishable from process 3 as the bottom position of the methyl group is distinct from the upper positions, whereas the upper positions are equivalent. **B.** In CH$_3$NH$_2$, all of the exchange processes result in the same molecular conformation and are indistinguishable.

such, fourth-order term for the case of indistinguishable molecular processes becomes:

$$\frac{\partial}{\partial t}\rho^{(4)} = \sum_{jk}\frac{1}{4}\left[\hat{A}_j,\left[\hat{A}_j^\dagger,\left[\hat{A}_k,\left[\hat{A}_k^\dagger,\rho\right]\right]\right]\right]\frac{T}{2\tau^2}$$
$$= \left(\sum_k \frac{\hat{\mathcal{L}}_k}{\tau_k}\right)^2 \rho \frac{T}{2} \quad (26)$$

Note that the $j$-index can be removed by raising the summation to a power.

Notice that in both **eq. 25** and **26**, the dissipator is proportional to the square of the Lindbladian, indicating that the fourth-order term encapsulates the probability of two exchange events. Furthermore, this term will naturally disappear as the step size $T \to dt$, because only one exchange event will be probable in the differential limit.

Following the same procedure as in deriving the fourth-order interactions, we may recast the entire Dyson expansion in terms of the exchange Lindbladian, which in the case of distinguishable exchange processes is

$$\frac{\partial}{\partial t}\rho = \left\{\sum_{n=0}^{\infty}\sum_k \left(\frac{\hat{\mathcal{L}}_k}{\tau_k}\right)^{n+1}\frac{1}{n!}\left(\frac{T}{2}\right)^n\right\}\rho, \quad (27)$$

and similarly for the case of distinguishable ensembles:

$$\frac{\partial}{\partial t}\rho = \left\{\sum_{n=0}^{\infty}\left(\sum_k \frac{\hat{\mathcal{L}}_k}{\tau_k}\right)^{n+1}\frac{1}{n!}\left(\frac{T}{2}\right)^n\right\}\rho \quad (28)$$

In either of these cases, the most challenging aspect of evaluating these equations is calculating the infinite powers of the Lindbladian, and is typically why perturbation theory is only extended to the few lowest-order terms. However, we previously noted when deriving the DMEx that these equations may be dramatically simplified if the dissipator, in this case the Lindbladian, obeys the property

$$\hat{\mathcal{L}}^{n+1}\rho = \hat{\mathcal{L}}^n\hat{\mathcal{L}}\rho = \gamma_n\hat{\mathcal{L}}\rho, \quad (29)$$

which states that $\hat{\mathcal{L}}\rho$ is an eigenfunction of $\hat{\mathcal{L}}^n$ with an eigenvalue $\gamma_n$, which is a scalar quantity. For indistinguishable exchange processes, $\hat{\mathcal{L}}$ is the $k$-sum of individual Lindbladians and for distinguishable exchange processes, $\hat{\mathcal{L}} = \hat{\mathcal{L}}_k$. In that case, all higher order Lindbladians may be written as being proportional to the lowest order term. If **eq. 29** is obeyed, then the infinite Lindblad series for exchange is given by:

$$\frac{\partial}{\partial t}\rho = \left\{\sum_k \frac{\hat{\mathcal{L}}_k}{\tau_k}\sum_{n=0}^{\infty}\frac{\gamma_n}{n!}\left(\frac{T}{2\tau_k}\right)^n\right\}\rho, \quad (30)$$

This result is identical for both indistinguishable and distinguishable exchange processes. Now, the infinite sum is proportional only to scalar quantities, and all higher order Lindbladian dissipators are simply proportional to $\hat{\mathcal{L}}_k$. We define the infinite sum as the exchange generating function

$$\sum_{n=0}^{\infty}\frac{\gamma_n}{n!}\left(\frac{T}{2\tau_k}\right)^n = \Gamma\left(\frac{T}{2\tau_k}\right), \quad (31)$$

which relates all higher order exchange interactions to the lowest-order interaction. Using this definition, we may write the exact Lindblad master equation for chemical exchange as:

$$\frac{\partial}{\partial t}\rho = \sum_k \frac{\hat{\mathcal{L}}_k\rho}{\tau_k}\Gamma\left(\frac{T}{2\tau_k}\right) \quad (32)$$

The only difference between the LMEx and the LME2 (c.f. **eq. 17**) is the exchange generating function, which is a scalar correction factor. In following, we will examine the systems that satisfy **eq. 29**.

## III. Generating exact Lindbladians

To construct an exact Lindblad master equation, one must determine the form of the scalar exchange generating function $\Gamma(T/2\tau_k)$. To do this, the condition set by **eq. 29** must be met and the series of $\gamma_n$ eigenvalues must be known. In the case of exchange between distinguishable exchange processes, one can calculate that the series $\gamma_n$ takes the form

$$\gamma_n = (-2)^n, \qquad (33)$$

which when used in **eq. 31** gives:

$$\Gamma\left(\frac{T}{2\tau_k}\right) = \sum_{n=0}^{\infty} \frac{(-2)^n}{n!}\left(\frac{T}{2\tau_k}\right)^n = \exp\left(\frac{-T}{\tau_k}\right) \qquad (34)$$

Using this result in **eq. 30** gives the LMEx for distinguishable exchange processes:

$$\frac{\partial}{\partial t}\rho = \sum_k \frac{\hat{\mathcal{L}}_k \rho}{\tau_k} \exp\left(\frac{-T}{\tau_k}\right) \qquad (35)$$

Note that the exchange generating function is an elementary, scalar function that is included at no additional computational cost. Casting the system in a basis where the set of exchange processes are uncorrelated effectively casts the entire system as a sum of two-site exchange processes (**Fig. 4**). In this case, one could readily motivate the form of the exchange generating function by accounting for the probability for any $n$-exchange event and using that to scale the likelihood of the system appearing to evolve as $\hat{\mathcal{L}}\rho$.

The case of indistinguishable exchange processes is slightly more complicated to derive the exchange generating function, as all transitions are coupled together. The different molecular conformations of these systems are often permutational isomers, and thus the operators defining transitions between the conformations may be described by permutation groups. The order of the group corresponds to the number of configurations in the system. In this case, system geometries that satisfy **eq. 29** for the combinations of transitions that form different permutation groups of order $f$, which can be done rapidly as it only requires the Fock-space. To validate **eq. 29**, one need only check to see if $\mathcal{L}^2\rho \propto \mathcal{L}\rho$, which immediately satisfy **eq. 29** for all powers of the Lindbladian. For example, the set of operators $\{\hat{A}_k\}$ that permit exchange between three configurations are:

$$\left\{\begin{pmatrix}0 & 1 & 0\\1 & 0 & 0\\0 & 0 & 1\end{pmatrix},\begin{pmatrix}1 & 0 & 0\\0 & 0 & 1\\0 & 1 & 0\end{pmatrix},\begin{pmatrix}0 & 0 & 1\\0 & 1 & 0\\1 & 0 & 0\end{pmatrix}\right\} \qquad (36)$$

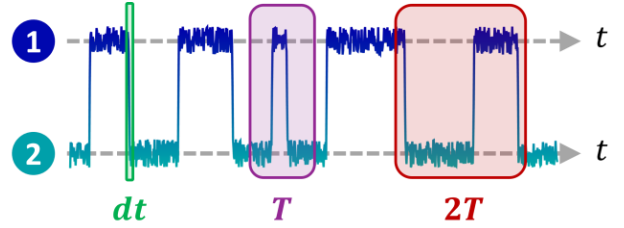

**Figure 4 | The exchange generating function accounts for higher-moments of chemical exchange.** Over the period $dt$, a molecule only has the opportunity to undergo a single exchange event. However, multiple exchange events are probable over a finite period of time $T$, which are accounted for by the exchange generating function. As $T$ increases, the probability of higher-order exchange terms appearing increases. Hence, as $T \to \infty$, the probability of the system appearing to undergo an exchange event of the form $\hat{\mathcal{L}}\rho$, the leading order term, asymptotically approaches zero.

These operators correspond to the Fock-space, which is only ever populated on the diagonal elements of the space. Using an initial arbitrary Fock density matrix of

$$\hat{\rho} = \begin{pmatrix}a & 0 & 0\\0 & b & 0\\0 & 0 & c\end{pmatrix}, \qquad (37)$$

one can evaluate the expression

$$\sum_{i \in k} \hat{\mathcal{L}}_i \left(\sum_{j \in k} \hat{\mathcal{L}}_j \hat{\rho}\right) = \gamma_1 \left(\sum_{j \in k} \hat{\mathcal{L}}_j \hat{\rho}\right) \qquad (38)$$

for when $\gamma_1$ is a constant, which relates the linear ($\hat{\mathcal{L}}$) and quadratic ($\hat{\mathcal{L}}^2$) Lindbladians. The indices $i \in k$ and $j \in k$ span permutations of the exchange operators that connect each configuration and, importantly, $i$ and $j$ must span the same permutational set of operators ($\{i\} = \{j\}$). The groups that satisfy **eq. 29** are ones where every configuration has a transition to every other configuration. In general, these groups can be recast into a pseudorotation group defined by a forward ($\hat{R}$) and backward ($\hat{R}^{-1}$) rotation of the Hilbert space, and are Abelian groups. In this case, the series $\gamma_n$ for a permutation group of order $h$ with $N = (h^2 - h)/2$ transitions is

$$\gamma_n = \left(\frac{-h}{N}\right)^n, \qquad (39)$$

which makes the LMEx for these groups:

$$\frac{\partial}{\partial t}\rho = \frac{1}{N}\sum_k^N \frac{\hat{\mathcal{L}}_k \rho}{\tau} \exp\left(\frac{-hT}{2N\tau}\right) \qquad (40)$$

As the rates for indistinguishable processes will be identical, we replace $\tau_k \to \tau$. Each of the distinguishable pathways of the previous case can simply be thought of a pseudorotation group of order 2, for which **eq. 39** predicts **eq. 33**.

Many systems cannot be represented by a pseudorotation group. The simplest of these systems is a linear (acyclic) 3-state system, which can be thought of as a cyclic 3-state system that belong to a permutation group of order three ($\mathcal{G}^3$), but missing two transitions, which belong to a permutation group of order two ($\mathcal{G}^2$). As such, this can be thought of as a system belonging to a $\mathcal{G}^3 - \mathcal{G}^2$ permutation group, which are strictly non-Abelian permutation groups and cannot be cast as a pseudorotation. In general, we can write equation of motion for the case of a $\mathcal{G}^h - \mathcal{G}^f$ system, where $h > f$ and there are $N$ and $M$ transitions in $\mathcal{G}^h$ and $\mathcal{G}^f$, respectively:

$$\frac{\partial}{\partial t}\rho = \frac{1}{N}\sum_k^N \frac{\hat{\mathcal{L}}_k \rho}{\tau}\exp\left(\frac{-hT}{2N\tau}\right)$$
$$-\frac{1}{N}\sum_{j \in k}^M \frac{\hat{\mathcal{L}}_j \rho}{\tau}\Gamma'\left(\frac{T}{2\tau}\right) \quad (41)$$

For the $\mathcal{G}^3 - \mathcal{G}^2$ example, $h = 3, N = 3, f = 2, M = 1$. We will refer to the $\mathcal{G}^h$ as the head group, which provides all potential transitions and determines the number of conformations in the system, and $\mathcal{G}^f$ as the non-Abelian forming group, which removes the appropriate transitions to generate the non-Abelian LMEx. Note that the terms corresponding to $\mathcal{G}^f$ in **eq. 41** are scaled to the number of transitions in the head group, which takes advantage of the linearity of the Lindblad equation. While $\mathcal{G}^f$ must be a pseudorotation group to satisfy **eq. 29**, the form of the generating function for $\mathcal{G}^f$, $\Gamma'(T/2\tau)$, will not simply be given by **eq. 39** and must satisfy a modified condition to generate an LMEx given by:

$$\left(\sum_k \hat{\mathcal{L}}_k \rho\right)^n - \left(\sum_{j \in k}\hat{\mathcal{L}}_j \rho\right)^n$$
$$= \gamma_n \left(\sum_k \hat{\mathcal{L}}_k \rho - \sum_{j \in k}\hat{\mathcal{L}}_j \rho\right) \quad (42)$$

We find that the set $\{j \in k\}$ transitions that satisfy **eq. 42** form another pseudorotation group. As such, this formulation of non-Abelian permutation groups must always be generated by the difference between two pseudorotation groups. Solving **eq. 42** for the series $\gamma_n$ for a generic $\mathcal{G}^h - \mathcal{G}^f$ system gives

$$\gamma_n = \left(\frac{-h}{N}\right)^n - \frac{h-f}{f}\left(\left(-\frac{h-f}{N}\right)^n - \left(\frac{-h}{N}\right)^n\right), \quad (43)$$

which leads to generating functions for the forming group $\mathcal{G}^f$, given by:

$$\Gamma'\left(\frac{T}{2\tau}\right) = \exp\left(\frac{-hT}{2N\tau}\right)$$
$$-\frac{h-f}{f}\left(1 - \exp\left(\frac{-fT}{2N\tau}\right)\right)\exp\left(\frac{-(h-f)T}{2N\tau}\right) \quad (44)$$

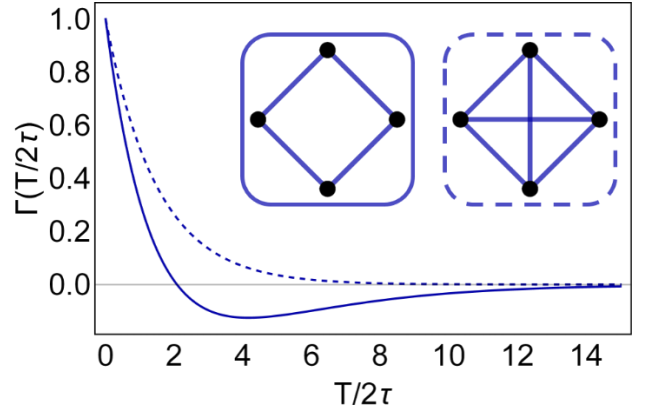

**Figure 5 | Generating functions for permutation groups of order 4.** $\Gamma'(T/2\tau)$ is shown for the case of $\mathcal{G}^4 - 2\mathcal{G}^2$ (solid) compared to the generating function of the $\mathcal{G}^4$ head group (dashed). The bipolar nature of the generating function for the non-Abelian forming group indicates that the population flux through the omitted transitions must change directions.

Even though this is more complicated than its counterpart in **eq. 40**, the generating function is still a relatively simple, scalar equation. Importantly, there can be multiple non-Abelian forming groups which each contribute a term as in **eq. 41** with the only restriction that they must span entirely separate transitions ($[\mathcal{G}^f, \mathcal{G}^{f'}] = 0$). The behavior of this generating function is also interesting, as it is bipolar (**Fig. 5**). This indicates that the flow of polarization will actually reverse $T/2\tau$ to account for the number of exchange events through the pathways that are in the head group but should be absent. Furthermore, we generally find that this generating function, along with the generating function for pseudorotation groups (**c.f. eq 40**) decay more slowly as the order of the head group increases. This can be interpreted as the system requiring more transitions, and thus more time, to reach the point where it cannot be described by the leading term $\hat{\mathcal{L}}\rho$, and rather described by a system that is randomly configured with respect to where it started. Finally, it should be mentioned that this example is not necessarily the only formulation that would permit unique non-Abelian permutation groups where the exchange processes are indistinguishable. However, it is an example for how one can develop scalar exchange generating functions.

### IV. Performance of exact Lindblad master equations

To highlight the importance of utilizing exact Lindblad master equations for chemical exchange, we will explore the convergence radius of the solution of the master equations presented here in comparison to the traditional master equation for chemical exchange in Lindblad form, **eq. 17**. Importantly, we will use a first order integration technique to emphasize the improvement in the solution convergence offered by the exchange generating function. Specifically, we will calculate the solution at a time $t + T$ to be:

$$\hat{\rho}(t+T) = \hat{U}\hat{\rho}(t)\hat{U}^\dagger + \frac{T}{\tau}\hat{\mathcal{L}}\rho(t)\Gamma\left(\frac{T}{2\tau}\right) \qquad (45)$$

For all of the following cases, we will compare the solutions to an LMEx simulation calculated with $\Delta t \ll \tau$, which we will denote as the ground truth simulation. Unless noted otherwise, $T/\tau = 1\%$ for the ground truth simulation, which limits the probability of even two exchange events (the fourth-order Dyson term) to be $0.01\%$ during this period. For systems that multiple exchange rates, this condition was set based on the fastest rate.

We will first examine the case of distinguishable molecular processes, for which we can use **eq. 35**. To emphasize the broad generality of this technique, we will simulate a technique at the forefront of magnetic resonance, Signal Amplification By Reversible Exchange[22-28], or SABRE (**Fig. 6A**). In this method, nuclear spin polarization is transferred from the singlet order of parahydrogen ($p$-H$_2$) to artificially induce large magnetic resonance signals that are orders of magnitude larger than in conventional magnetic resonance. This system exists in a regime where the dominant couplings and resonance frequency differences are often on the order of the exchange rates of the system. Furthermore, nonlinear effects dominate this system and are dictated by the chemical exchange, making this system a sensitive reporter of the performance of the simulation. Full details of the physical model for SABRE have been reported elsewhere.

For the case of distinguishable exchange processes, we find the convergence radius of the LMEx equation can be up to an order of magnitude larger than that of the traditional, second-order exchange term in Lindblad form (**Fig. 6B**). While the traditional formulation exceeds a 1% error when $T/\tau \approx 2.5\%$, the LMEx formulation does not exceed this error until $T/\tau \approx 38.5\%$, which represents an approximately 15-fold increase in the convergence radius of this system for no additional computational cost.

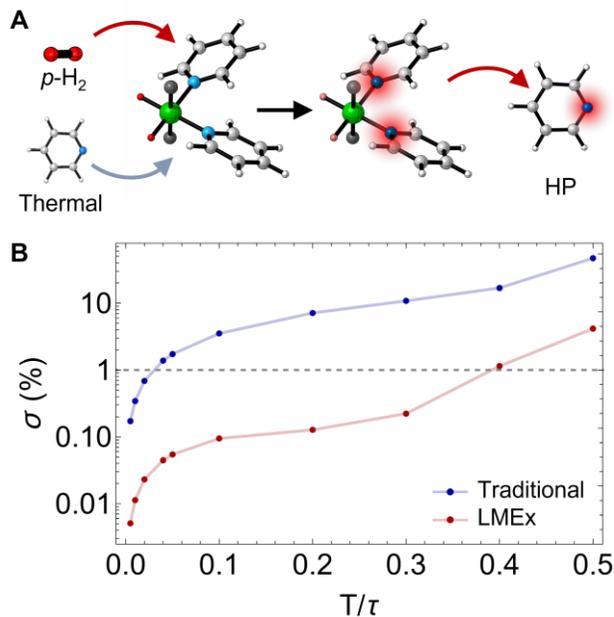

**Figure 6 | Modeling distinguishable exchange processes within the LMEx formalism. A.** Signal Amplification By Reversible Exchange (SABRE) is a hyperpolarization technique at the forefront of magnetic resonance. Nuclear spin polarization is distilled from the singlet order of parahydrogen ($p$-H$_2$) during reversible exchange interactions with an organometallic catalyst (green). **B.** The root-mean-squared deviation (RMSD, $\sigma$) between the traditional (blue) and LMEx (red) solutions. We find superior convergence of the solution at no additional computational cost. The $\sigma = 1\%$ line is demarcated to guide the eye, representing 99% solution convergence.

There are many coupled exchange mechanisms in the SABRE system, namely the exchange of parahydrogen and the target ligand with the catalyst. This would appear to violate the restriction that was imposed that exchange processes were only self-correlated. However, we circumvent this by formulating the exchange pathways in a way where the pathways are uncoupled. The parahydrogen exchange ($\hat{F}_k$) and target ligand exchange ($\hat{F}_j$) pathways are transformed to give the joint probability of either ligand exchange with parahydrogen exchange ($\hat{F}_j + \hat{F}_k$) or ligand exchange without parahydrogen exchange ($\hat{F}_j - \hat{F}_k$). Then, the two new pathways are effectively orthogonal and satisfy self-correlation requirement. In total, there are 13 different exchange processes that are all accounted for within the scope of this LMEx master equation, highlighting the flexibility of this framework for improving the performance of chemical exchange simulations.

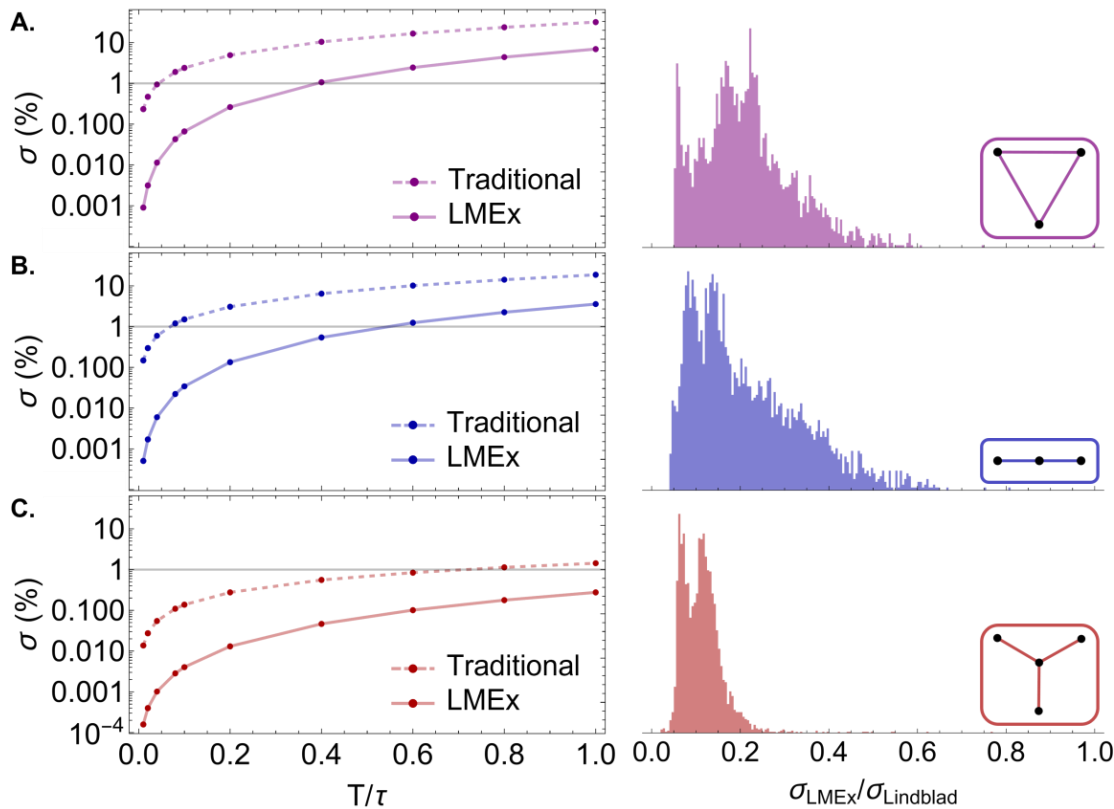

**Figure 7 | Exploring performance of LMEx models for systems with indistinguishable exchange processes.** The convergence of the solutions (left) were all calculated using the same coherent interaction parameter set, for consistency. Furthermore, the system parameters were iterated 3,200 times at $T/\tau = 20\%$ to examine the robustness of this performance. The histograms of the ratio between the LMEx and traditional errors are shown to the right. **A.** The $\mathcal{G}^3$ system can be cast as a pseudorotation and has a relatively simple exchange generating function, given by a single exponential factor $\Gamma(T/\tau) = exp(-T/\tau)$. For this case, we see that the solution is has a convergence radius that is approximately a factor of 10 larger than the traditional solution and that this is robust as the system parameters are varied. However, the **B.** $\mathcal{G}^3 - \mathcal{G}^2$ and **C.** $\mathcal{G}^4 - \mathcal{G}^3$ systems have more complicated generating functions that have to be constructed according to eq. 41 and 44. Despite this, they exhibit solution convergences that are on the order of or even exceed the $\mathcal{G}^3$ system and have similar robustness to changes in the coherent interactions.

In addition to the case of distinguishable exchange processes, there are a vast scope of systems that have indistinguishable exchange processes. Broadly, these are systems with permutation symmetry, for which an extensive amount of work has been devoted to calculating the dynamic NMR spectra for such systems. Here, we examine the case of a four spin-1/2 system that has permutation group symmetry belonging to a $\mathcal{G}^3$ pseudorotation group, a $\mathcal{G}^3 - \mathcal{G}^2$ permutation group, as well as a $\mathcal{G}^4 - \mathcal{G}^3$ permutation group, using the notation that we described previously. While the first of these may be cast as an Abelian pseudorotation group, the latter two cases cannot and are incompatible with our previous superoperator-based methods for chemical exchange. In addition to examining the convergence of the solutions for these systems, we are able to iterate over the parameters that define the coherent interactions (resonance frequencies and couplings) to assess the robustness of the improvement that the LMEx provides over the traditional implementation.

We find that an appropriately formulated exchange generating function can yield a vast improvement in the convergence radius of the solution for no additional computational cost, even for the non-Abelian permutation groups that were tested here (**Fig. 7**). In every case, the LMEx exhibits a robust improvement in the solution convergence that is approximately an order of magnitude. Furthermore, we find that the improvement that is obtained from the LMEx formulation is robust to variations in the system parameters. Using the case $T/\tau = 20\%$, we find that $\langle \sigma_{LMEx}/\sigma_{Lindblad} \rangle = (21 \pm 12)\%$ for the $\mathcal{G}^3$ system, at $\langle \sigma_{LMEx}/\sigma_{Lindblad} \rangle = (22 \pm 13)\%$ for the $\mathcal{G}^3 - \mathcal{G}^2$ system, and at $\langle \sigma_{LMEx}/\sigma_{Lindblad} \rangle = (11.4 \pm 5.3)\%$ for the $\mathcal{G}^4 - \mathcal{G}^3$ system. Out of the 9,600 different system parameterizations that were tested over these three cases, there was only a single instance where the LMEx did not yield a superior convergence over the traditional master equation for chemical exchange.

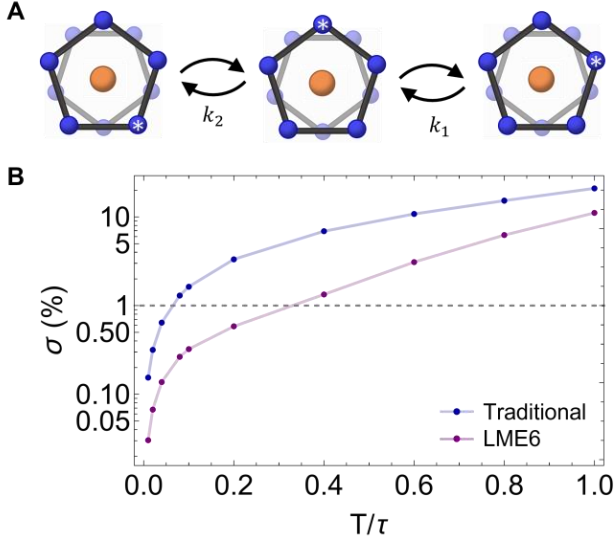

**Figure 8 | Improving performances of systems that do not yield scalar generating functions. A.** The fivefold rotation of permethylferrocene undergoes single jump ($k_1$) and double jump ($k_2$) rotations about the symmetry axis of this system. These pathways are distinguishable from one another, but each have five indistinguishable exchange processes that form the group of exchange operators. **B.** We find that calculating the first three non-zero terms of the Dyson series (LME6) offers a relatively inexpensive route to improving the convergence radius of the solution. Here we show a 5.4× improvement in the convergence radius at a cost of an additional 60% computational time, on average.

Both the cases for distinguishable and indistinguishable exchange processes show that accounting for all higher moments of the exchange interaction in the equation of motion yields a vastly improved convergence radius of the solution. We regularly found instances where the convergence radius had increased by an order of magnitude and that this was robust to changes in the system parameters. However, there are limits to which systems will have scalar exchange generating functions. One such example is rotation of five-fold symmetric systems as is observed in permethylferrocene[29] (**Fig. 8**). This system has two distinguishable sets of processes that are comprised of single jump ($2\pi/5$) or double jump ($4\pi/5$) rotations about the C5 axis of this system. While these exchange processes are distinguishable from one another, there are five individual indistinguishable processes that are required to fully describe this system. The particular five-fold cyclic geometry that is discussed in this case yields no scalar exchange generating function, at least to the point that has developed in this work. However, in rare cases such as this, the Fock-space representation of this problem offers a convenient solution to calculate the powers of the Lindbladian, and the rest of the Dyson series is already summarized in its most general form for chemical exchange in **eq. 28**. This is most efficient to compute symbolically, such that various symbolic variables may be replaced with the appropriate density matrix during the calculation. Thus, it is possible to simply calculate the first few terms of the series as a last resort for special cases.

Here, we demonstrate the improved convergence radius of an LME6 simulation, which utilizes the first three terms of **eq. 28** (to sixth order in the Dyson series), yields an approximately 5.4-fold improvement in the convergence radius compared to the traditional master equation and only requires approximately 60% more computational time to evaluate, which ultimately yields superior performance of the solution. The $m^\text{th}$ order Lindbladian for the single ($\hat{\mathcal{L}}_1^{(m)}$) and double ($\hat{\mathcal{L}}_2^{(m)}$) jump exchange pathways applied to the site $n$ for this case are given by:

$$\left[\hat{\mathcal{L}}_1^{(1)}\hat{\rho}\right]_n = \frac{1}{5}(\hat{\rho}_{n-1} + \hat{\rho}_{n+1} - 2\hat{\rho}_n)$$

$$\left[\hat{\mathcal{L}}_1^{(2)}\hat{\rho}\right]_n = \frac{1}{25}(6\hat{\rho}_n - 4(\hat{\rho}_{n-1} + \hat{\rho}_{n+1}) + \hat{\rho}_{n-2} + \hat{\rho}_{n+2})$$

$$\left[\hat{\mathcal{L}}_1^{(3)}\hat{\rho}\right]_n = \frac{1}{25}(3(\hat{\rho}_{n-1} + \hat{\rho}_{n+1}) - \hat{\rho}_{n-2} - \hat{\rho}_{n+2} - 4\hat{\rho}_n)$$

$$\left[\hat{\mathcal{L}}_2^{(1)}\hat{\rho}\right]_n = \frac{1}{5}(\hat{\rho}_{n-2} + \hat{\rho}_{n+2} - 2\hat{\rho}_n)$$

$$\left[\hat{\mathcal{L}}_2^{(2)}\hat{\rho}\right]_n = \frac{1}{25}(6\hat{\rho}_n - 4(\hat{\rho}_{n-2} + \hat{\rho}_{n+2}) + \hat{\rho}_{n-1} + \hat{\rho}_{n+1})$$

$$\left[\hat{\mathcal{L}}_2^{(3)}\hat{\rho}\right]_n = \frac{1}{25}(3(\hat{\rho}_{n-2} + \hat{\rho}_{n+2}) - \hat{\rho}_{n-1} - \hat{\rho}_{n+1} - 4\hat{\rho}_n)$$

Additionally, there is no significant difference in the performance of the LME6 solution in comparison to a solution that utilizes the first 40 terms of the Dyson expansion. While it is not ideal to have to result to brute-force evaluation of **eq. 28**, here we demonstrate that in the rare cases that are not already described here that the first few terms of the Dyson series can significantly improve computational performance at modest cost.

In this section, we have examined the performance of various exact Lindblad master equations for chemical exchange. To summarize the various dissipators that should be used for the type of exchange processes, the various cases are in Table 1 below along with references to the equations in which they were derived. All distinguishable exchange processes may be treated within this formalism, and we have introduced various methods to handle indistinguishable exchange processes in an exact fashion. Finally, we have shown that the Dyson series may be brute-force evaluated up to order $K$ and can still generate significant computational improvements.

| Process Type | Dissipator | Eq. |
|---|---|---|
| Distinguishable (all types) | $\sum_k \frac{\hat{L}_k \rho}{\tau_k} \exp\left(\frac{-T}{\tau_k}\right)$ | (35) |
| Indistinguishable pseudorotation | $\frac{1}{N} \sum_k^N \frac{\hat{L}_k \rho}{\tau} \exp\left(\frac{-hT}{2N\tau}\right)$ | (40) |
| Indistinguishable non-Abelian* permutation | $\frac{1}{N} \sum_k^N \frac{\hat{L}_k \rho}{\tau} \exp\left(\frac{-hT}{2N\tau}\right)$ $-\frac{1}{N} \sum_{j \in k}^M \frac{\hat{L}_j \rho}{\tau} \Gamma'\left(\frac{T}{2\tau}\right)$ | (41) & (44) |
| Other indistinguishable processes; order $K$ correction | $\sum_{n=0}^K \left(\sum_k \frac{\hat{L}_k}{\tau_k}\right)^{n+1} \frac{1}{n!} \left(\frac{T}{2}\right)^n$ | (28) |

**Table 1|** Exact Lindblad equations derived in this work, where $h$ corresponds to the number of sites and $N = (h^2 - h)/2$.

## V. Conclusions

Here, we have generalized and extended the exact master equation treatment of chemical exchange phenomena within the Lindblad formalism. We have developed methods for identifying exact master equations for a variety of systems and have presented solutions for any system with distinguishable exchange processes and a vast scope of systems with indistinguishable exchange processes. We have found that the exact Lindblad master equation approach generates an approximately order of magnitude larger convergence radius for the variety of systems that we studied at no additional computational cost. Furthermore, this improvement is robust to changes in the system-specific parameters. This result is trivial to implement into any existing chemical exchange simulation and has shown to vastly improve the convergence of the solution.

## Conflicts of interest

There are no conflicts to declare.

## Acknowledgements

This work was supported by the National Science Foundation grant CHE-2003109.